\documentclass[twocolumn,pra]{revtex4}
\usepackage{graphicx}
\usepackage{amsmath,amssymb}
\newcommand{\ket}[1]{|{#1}\rangle}			
\newcommand{\bra}[1]{\langle{#1}|}

\begin{document}
\title{Blind post-processing for the unbalanced BB84}

\author{Satoshi Sunohara$^{1}$}
\email{sunohara@qi.mp.es.osaka-u.ac.jp}
\author{Kiyoshi Tamaki$^{2,3}$}
\email{tamaki.kiyoshi@lab.ntt.co.jp}
\author{Nobuyuki Imoto$^{1}$}
\email{imoto@mp.es.osaka-u.ac.jp}
\affiliation{
$^{1}$Division of Materials Physics, Graduate School of Engineering Science,Osaka University,\\
1-3 Machikaneyama, Toyonaka, Osaka 560-8531, Japan\\ 
$^{2}$NTT Basic Research Laboratories, NTT Corporation,\\
3-1,Morinosato Wakamiya Atsugi-Shi, Kanagawa, 243-0198, Japan\\ 
$^{3}$National Institute of Information and Communications Technology, \\
4-2-1 Nukui-Kita, Koganei, Tokyo 184-8795, Japan
}

\begin{abstract}
For the realization of quantum key distribution, it is important to investigate its security based on a mathematical model that captures properties of the actual devices used by the legitimate users. Recently, Ferenczi, {\it et. al.} (Phys. Rev. A {\bf 86} 042327 (2012)) pointed out potential influences that the losses in phase modulators and/or the unbalance in the transmission rate of beam splitters may have on the security of the phase-encoded BB84 and analyzed the security of this scheme, which is called the unbalanced BB84. In this paper, we ask whether blindly applying the post-processing of the balanced BB84 to the unbalanced BB84 would lead to an insecure key or not, and we conclude that we can safely distill a secure key even with this post-processing. It follows from our proof that as long as the unbalances are basis-independent, our conclusion holds even if the unbalances are unknown and fluctuate in time.
\end{abstract}

\pacs{03.67.Dd, 03.67.-a}


\maketitle

\section{Introduction}
Quantum key distribution (QKD) is a protocol to share the secret key between two authenticate parties (Alice and Bob) with negligible leakage of its information to an eavesdropper (Eve). The advantage of employing QKD is that it can achieve the unconditional security, which is the security against any possible attack allowed by the law of quantum mechanics under some assumptions on the devices used by Alice and Bob.

The first QKD protocol was proposed by Bennett and Brassard at 1984 \cite{BB84} (the protocol is called BB84 protocol). Since this proposal, many works have been devoted to prove the unconditional security \cite{Mayers, Lo and Chau, ILM, Shor and Preskill, Nonorthogonal, Koashi}, and some works take into account practical imperfections of the devices used by Alice and Bob \cite{Koashi and Preskill, GLLP, FTQLM}. It is important that a mathematical model of Alice and Bob's devices is needed to prove the security of a QKD protocol, and thus the model should reflect the actual imperfections of the devices for the realization.

In this paper, we consider the effect of the losses in phase modulators and the unbalance in the transmission rate of beam splitters in the phase encoded BB84. This practical imperfection is firstly pointed out by Ferenczi, {\it et. al.} \cite{Original}, and this  
scheme is referred to as the unbalanced BB84. Since the actual phase modulators have inevitable losses and the transmission rate of actual beam splitters cannot be exactly 50\%, it is important to analyze the security of the protocol in order to fit the theory to the actual situation. The security of this protocol has been analyzed by Ferenczi, {\it et. al.} \cite{Original} based on the security proof \cite{Devetak and Winter, KGR}. In their proof, however, it is not clear whether blindly applying the post-processing for the standard (balanced) BB84 would lead to an insecure key or not. 

The purpose of this paper is to provide the unconditional security proof of the unbalanced BB84 by showing that any security proof
of the balanced BB84 where Eve is allowed to distribute Alice and Bob a basis-independent state, for instance the security proof based on complementary scenario \cite{Koashi} or Shor-Preskill type proof \cite{Shor and Preskill}, can be directly applied to the unbalanced BB84. This means that we can safely perform the data processing for the key distillation as if there were no unbalance and the unbalance only changes experimental data. Moreover, a natural consequence of our security proof is that as long as the unbalances are basis-independent, our conclusion holds even if the unbalance in Alice and Bob is unknown and fluctuate in time. 

In order to see the performance of the unbalanced BB84, we simulate the resulting key generation rate as a function of the distance between Alice and Bob. Following the work by Ferenczi, {\it et. al.}, we consider two cases: the first case is that we employ the unbalanced BB84 as it is and the second one is that we apply additional attenuations to Alice and Bob's devices in order to balance the intensities of the double pulses and to eliminate the effect of the unbalance in Bob's measurement. We call the second case as the BB84 with the hardware fix or the hardware fix scenario, and this case is essentially the same as standard BB84 with additional losses. By simulating the key generation rates for the two cases, we have obtained almost the same threshold distance of the key generation in \cite{Original} and confirmed that the key rate of the BB84 with the hardware fix is lower than that of the unbalanced BB84 \cite{Original}. 

The organization of this paper is as follows, In Sec. \ref{Protocol}, we briefly review the protocol of the unbalanced BB84. In Sec. \ref{Security proof}, we first briefly review the security proof of the balanced BB84 and prove the unconditional security of the unbalanced BB84. In Sec. \ref{Simulation}, by assuming the use of the decoy state \cite{Decoy}, we simulate the key generation rates of the unbalanced BB84 and the BB84 with the hardware fix. The key rates are plotted as a function of the distance between Alice and Bob by taking experimental data from GYS experiment \cite{GYS}. Finally, we summarize this paper in Sec. \ref{summary}.

\section{Unbalanced BB84}\label{Protocol}
\begin{figure}[tbh]
\begin{center}
 \includegraphics[scale=0.35]{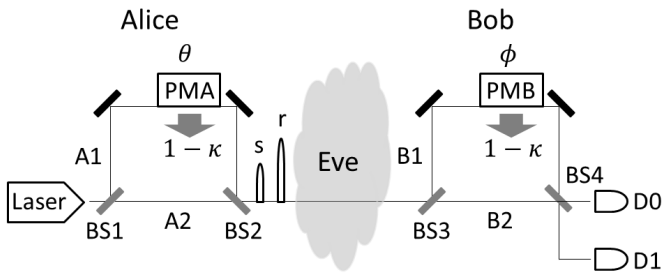}
\end{center}
 \caption{Description of our model of the unbalanced BB84 protocol. PMA and PMB are the lossy phase modulators with the transmission rate of $\kappa$. Note that our proof is valid even if the unbalances in Alice and Bob's sides are different and unknown.}
 \label{Composition}
\end{figure}

In this section, we introduce the unbalanced BB84.
The experimental setup of the unbalanced BB84 is depicted in FIG. \ref{Composition}.
The unbalanced BB84 is the standard phase encoded BB84 protocol using phase randomized weak coherent pulses where the intensities of the double pulses are not the same
because of the imperfections of the phase modulators and/or the beam splitters.
For the simplicity of the discussion, we assume that the phase
modulators possessed by Alice and Bob have the same transmission
rate $\kappa$ and we do not explicitly consider the unbalance of the beam 
splitters, i.e., all the relevant beam splitters are assumed to have 50\% of 
transmission rate, but the generalization of our analysis is trivial.

In Alice's side, a phase randomized coherent pulse from her laser source splits into two arms A1 and A2 by a balanced beam splitter BS1. Alice randomly applies phase modulation $\theta=\{0,\pi/2,\pi,3\pi/2\}$ to the pulse passing A1 by the imperfect phase modulator PMA. Here, the phase $\theta=\{0,\pi\}\ (\{\pi/2,3\pi/2\})$ is defined as the bit value $\{0,1\}$ in $X\ (Y)$ basis (for the later convenience, we use $W=X,Y$ to refer the basis). The pair of the pluses come from A1 and A2 is sent to Bob via another balanced beam splitter BS2. Because of the imperfection of PMA, the state of the pulses sent by Alice becomes
${\ket{e^{i(\zeta+\theta)}\sqrt{\kappa \alpha}}}_s{\ket{e^{i\zeta}\sqrt{\alpha}}}_r$, where subscripts s and r respectively denote the signal pulse passed through A1 and reference pulse passed through A2, and $\zeta$ is the random phase chosen between 0 and $2\pi$.

In Bob's side, the optical length difference between B1 and B2 is adjusted to the same as the one between A1 and A2. Thanks to his interferometer, the pair of the incoming pulses is finally separated into three pulses which can be distinguished by the detection time of the detectors, and we only consider the pulses arriving at the intermediate time.
Bob applies the phase modulation $\phi$ randomly chosen from $\{0, \pi/2\}$ to the pulse passing through B1 by the imperfect phase modulator PMB. Here, the phase modulation $0\ (\pi/2)$ is defined as $X\ (Y)$ basis in Bob's measurement.
As the result of Bob's phase modulation, the state of the pulses arriving at the beam splitter BS4 becomes
${\ket{e^{i(\zeta+\theta)}\sqrt{\kappa \beta}}}_s{\ket{e^{i(\zeta+\phi)}\sqrt{\kappa \beta}}}_r$, and state at the detector D0\ (D1) becomes $\ket{e^{i\zeta}\frac{e^{i\phi}+e^{i\theta}}{\sqrt{2}}\sqrt{\kappa \beta}}$ $\left(\ket{e^{i\zeta}\frac{e^{i\phi}-e^{i\theta}}{\sqrt{2}}\sqrt{\kappa \beta}}\right)$.
Bob records the bit value 0\ (1) when D0\ (D1) clicks, and Bob randomly assigns a random bit to the double click event if the double click event occurs due to noises such as the dark counting of the detectors or misalignment. After Bob's measurement, Bob broadcasts his basis, and Alice and Bob keep the data with the bases matched. One can see that Bob can obtain the same bit value with Alice when there are no noises.

We note that we make the following assumption on Bob's detection device: the POVM element $\hat{F}_W^{(f)}$ corresponding to the failure detection of the bit value in $W$ basis is basis-independent, i.e.,  
\begin{align}
\hat{F}_X^{(f)}=\hat{F}_Y^{(f)}\,.\label{Basis Independent Failure}
\end{align}
Therefore, we consider Bob's $W$ basis measurement is constructed by a basis-independent filter $\hat{F}_{\rm Bob}$ followed by Bob's two-outcome, i.e., Bob’s bit value, $W$ basis measurement $\hat{M}_W$. We note that the squash model \cite{GLLP,Tsurumaru and Tamaki,BML} is not necessary in our proof.

\begin{figure}[bth]
\begin{center}
 \includegraphics[scale=0.40]{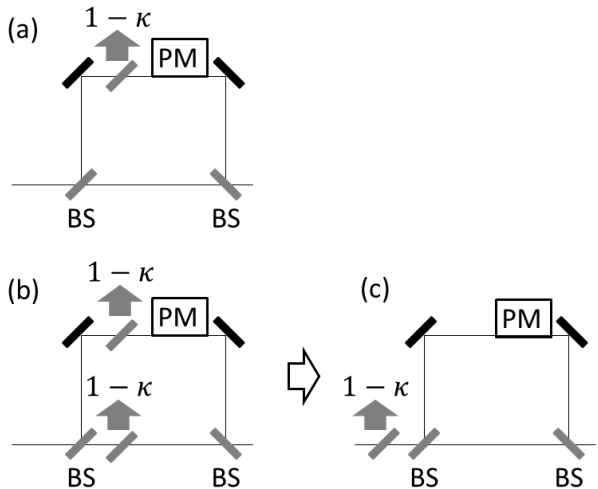}
\end{center}
 \caption{
(a) is the interferometer in the unbalanced BB84. (b) is the interferometer for the BB84 with the hardware fix. In (a) and (b), the loss in the phase modulator is modeled by the additional beam splitters with the reflection rate of $1-\kappa$ preceded by the lossless phase modulator (PM). (b) is equivalent to (c).
}
 \label{Fix}
\end{figure}

We remark that since each of the signal and reference pulses of our interest passes through the phase modulator only once, we do not need to equalize the 
intensities of the pulses in order to suppress the bit errors. However, just for the comparison, we consider
the case where we fix the unbalance by implementing the beam splitters with the transmission rate of $\kappa$
 to the paths of A2 and B2 (see also FIG. \ref{Fix}). We call this scenario as the BB84 with hardware fix or the hardware fix scenario, and we
can regard this case as the ideal BB84 with the additional attenuation. We will compare the key generation rates of both of the cases by the key generation simulations in Sec. \ref{Simulation}, and we confirm that the unbalanced BB84 protocol has larger key generation rate than the BB84 with the hardware fix.

\section{Security proof}\label{Security proof}
In this section, we prove the unconditional security of the unbalanced BB84 protocol based on complementarity scenario \cite{Koashi}. We consider only the single-photon part ($n=1$) since we may not be able to or may have very little chance to generate the key from multi-photon part due to the so-called photon-number splitting (PNS) attacks \cite{HIGM}. In order to treat the photon number separately, we apply the argument by GLLP \cite{GLLP} or by Koashi \cite{KoashiPractical}. For the simplicity of the discussion, we assume that the key is generated from the states of $X$ basis, and therefore the states of $Y$ basis are only spent in the parameter estimation, i.e., the so-called phase error estimation.

\subsection{The state of Alice and the brief review of the security proof of the balanced BB84}
Before the proof, we would like to discuss Alice's source. Thanks to the phase randomization, the density matrix of the pulses sent by Alice becomes the mixture of the eigenstates of the photon number $n$ (we rewrite the eigenstates as $\ket{a_W^{(n)}}_B$) as follows
\begin{align}
{\int_{0}^{2\pi}}\frac{d\zeta}{2\pi} \hat{P}( {\ket{e^{i(\zeta+\theta)}\sqrt{\kappa \alpha}}}_s{\ket{e^{i\zeta}\sqrt{\alpha}}}_r)=\sum_{n=0}^\infty p^{(n)}\hat{P}(\ket{a_W^{(n)}}_B)\label{Mixture}
\end{align}
where $W \in \{X,Y\}$ and $a \in \{0,1\}$ are the basis and bit value Alice chooses for sending the pulses ($W$ and $a$ in right-hand side corresponds to $\theta$ in the left-hand side), $\hat{P}$ is defined as $\hat{P}(\ket{\psi})=\ket{\psi}\bra{\psi}$, and the subscript $B$ represents the system to be sent to Bob.
The single-photon part ($n=1$) of $\ket{a_W^{(n)}}_B$ is described as $\ket{a_X^{(1)}}_B=\frac{1}{\sqrt{1+\kappa}}(\ket{0_Z}_B+(-1)^a\sqrt{\kappa}\ket{1_Z}_B)$ and
$\ket{a_Y^{(1)}}_B=\frac{1}{\sqrt{1+\kappa}}(e^{-i\frac{\pi}{4}}\ket{0_Z}_B+(-1)^a\sqrt{\kappa}e^{i\frac{\pi}{4}}\ket{1_Z}_B)$,
where we define $\ket{0_Z}_B={\ket{0}}_s{\ket{1}}_r$ and $\ket{1_Z}_B={\ket{1}}_s{\ket{0}}_r$.
For the later convenience, we define the relationships of the eigenstates of qubit states as follows:  $\ket{j_X}=\frac{1}{\sqrt{2}}(\ket{0_Z}+(-1)^j \ket{1_Z})$ and $\ket{j_Y}=\frac{1}{\sqrt{2}}(e^{-i\frac{\pi}{4}}\ket{0_Z}+(-1)^j e^{i\frac{\pi}{4}}\ket{1_Z})$ where $j=0,1$.

In the following two paragraphs, we briefly review the essential point of the security proof of the {\it balanced} BB84. In GLLP argument \cite{GLLP} or the argument by Koashi in \cite{KoashiPractical}, we ask whether the privacy amplification succeeds if it is applied only to the qubits 
associated with the single-photon emission, and it is shown that we can generate the secret key with asymptotic key generation rate $R$ as follow \cite{comment}. 
\begin{align}
R=-\gamma_Xf(E_X)h(E_X)+\gamma_X^{(1)}[1-h(e_{Y}^{'(1)})]\,.
\end{align}
Here, $e_{Y}^{'(1)}$ is $Y$ basis fictitious bit error rate that would have been obtained if Alice had sent a single-photon in $X$ basis and Alice and Bob had employed $Y$ basis for the measurement, $E_X$ is the bit error rate from the $X$ basis measurements when Alice sends a pulse in $X$ basis, $\gamma_X$ is the rate of Bob's detection in $X$ basis ($\gamma_X^{(1)}$ is the part of $\gamma_X$ where Alice sends a single photon), $f(E_X)$ is the inefficiency of the error correcting code, and $h(x)= -x \log_2x-(1-x) \log_2(1-x)$. 

Note that what the actual experiment gives Alice and Bob is the bit error rate in $Y$ basis $E_Y$ when Bob chooses $Y$ basis and Alice sends a pulse in $Y$ basis rather 
than a single-photon in $X$ basis. By using GLLP argument or combining GLLP argument with the decoy state idea \cite{Decoy}, we can estimate the lower bound of $e_{Y}^{(1)}$, which is the contribution from Alice's single-photon emission in $E_{Y}$, but this quantity is different from $e_{Y}^{'(1)}$. In the {\it balanced} BB84, it turns out that these quantities match ($e_{Y}^{(1)}=e_{Y}^{'(1)}$) \cite{Lo and Chau,Shor and Preskill,Koashi}. The most important point to derive $e_{Y}^{(1)}=e_{Y}^{'(1)}$ in the balanced BB84 is the basis independence of the state Alice prepares and of Bob's detection Eq. (1). Thanks to this independence, Eve cannot behave differently between $X$ and $Y$ basis so that we have $e_{Y}^{(1)}=e_{Y}^{'(1)}$ \cite{ILM,GLLP,Decoy,KoashiPractical}. In what follows, we prove that $e_{Y}^{(1)}=e_{Y}^{'(1)}$ is also the case for the unbalanced BB84, which means that we can perform the data processing in the unbalanced BB84 as if there were no unbalance. 

\subsection{Security proof of the unbalanced BB84}

For the security proof, we first consider a virtual protocol where we change the method to determine Alice's bit value $a$. In the virtual protocol, she firstly generates $\ket{\Psi_W^{+ (n)}}_{AB}$ with the probability $p^{(n)}$ 
\begin{align}
\ket{\Psi_W^{+ (n)}}_{AB}=\frac{1}{\sqrt{2}} 
(\ket{0_W}_A\ket{0_W^{(n)}}_B-\ket{1_W}_A\ket{1_W^{(n)}}_B)
\end{align}
where the subscripts $A$ and $B$ respectively denote Alice's and Bob's system, which is mathematically expressed by Hilbert space $H_A\bigotimes H_B$.
After generating the state, she conducts $W$ basis measurement on her qubit, and she records the measurement outcome as her bit value $a$ in $W$ basis. Finally, she sends the system $B$ to Bob. We can confirm the state sent to Bob is equivalent to the actual protocol.
In the virtual protocol, it does not matter if Alice delays her measurement after sending the state, and hence we assume this delay hereafter. 

From the definition of $\ket{\Psi_X^{+ (1)}}_{AB}$ and $\ket{\Psi_Y^{+ (1)}}_{AB}$, one can easily confirm that 
$\ket{\Psi_X^{+ (1)}}_{AB}=\ket{\Psi_Y^{+ (1)}}_{AB}$, i.e., the state of the single-photon in the virtual protocol is
basis-independent. By combining this independence with the basis-independence of Bob's measurement, we have $e_{Y}^{(1)}=e_{Y}^{'(1)}$. This ends the proof.

\begin{figure}[bht]
\begin{center}
 \includegraphics[scale=0.35]{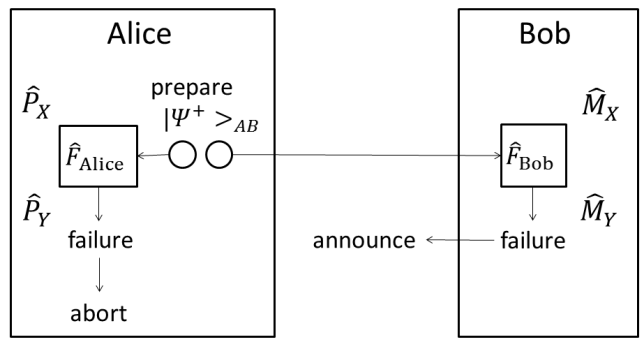}
\end{center}
 \caption{
Description of the virtual protocol for the single-photon part in our alternative proof. Alice (Bob) conducts the two outcome measurement of $W$ basis which consists of $\hat{P}_W$ ($\hat{M}_W$) following the post-selection by each side of the filters. Note that the filtering operations ${\hat F}_{\rm Alice}$ and ${\hat F}_{\rm Bob}$ are basis-independent, and as a result we can safely apply the security proof of the standard BB84 to the filtered state.
}
 \label{Virtual}
\end{figure}

As an alternative proof (also see FIG. \ref{Virtual}), we furthermore modify the single-photon part of the virtual protocol that is equivalent to the single photon part of the unbalanced BB84 protocol. Let $\hat{F}_{\rm Alice}=\sqrt{\kappa}\ket{0_Z}_{A}\bra{0_Z}+\ket{1_Z}_{A}\bra{1_Z}$ be Kraus operator corresponding to the successful event of her filtering operation. In the modified protocol, Alice first prepares a basis-independent joint state $\ket{\Psi^{+}}_{AB}=\frac{1}{\sqrt{2}}(\ket{0_Z}_A\ket{1_Z}_B+\ket{1_Z}_A\ket{0_Z}_B)$, performs the filtering operation, and then she keeps only the successfully filtered state. We can confirm that the filtered state satisfies $\hat{F}_{\rm Alice}\otimes{\hat \openone}_{B}\ket{\Psi^{+}}_{AB}\propto \ket{\Psi_X^{+ (1)}}_{AB}=\ket{\Psi_Y^{+ (1)}}_{AB}$, which means that the post-selected state is also basis-independent. This ends the alternative proof.

In summary, we can prove the security of the unbalanced BB84 protocol by directly confirming the basis-independence
$\ket{\Psi_X^{+ (1)}}_{AB}=\ket{\Psi_Y^{+ (1)}}_{AB}$ together with the basis-independence of Bob's measurement. Alternatively, we consider the virtual protocol for the single-photon part, which is constructed by the preparation of the Bell state $\ket{\Psi^{+}}_{AB}=\frac{1}{\sqrt{2}}(\ket{0_Z}_A\ket{1_Z}_B+\ket{1_Z}_A\ket{0_Z}_B)$ followed by each side of basis-independent filtering operation with $\hat{F}_{\rm Alice}$ and $\hat{F}_{\rm Bob}$, the post-selection, and each side of the two outcome measurements of $W$ basis. The two outcome measurements consist of $\hat{M}_W$ in Bob's side and projective $W$ basis qubit measurement $\hat{P}_W$ in Alice's side (depicted in FIG. \ref{Virtual}). This virtual protocol is composed only by basis-independent operations so that we have $e_{Y}^{(1)}=e_{Y}^{'(1)}$. 

From the above discussion, we conclude that the key generation rate of the unbalanced BB84 protocol is written as
\begin{align}
R=-\gamma_Xf(E_X)h(E_X)+\gamma_X^{(1)}[1-h(e_{Y}^{(1)})]\,.
\end{align}
This formula is completely equivalent to the rate of the balanced BB84 \cite{GLLP,Decoy,KoashiPractical}. Therefore, we can perform the data processing in the unbalanced BB84 as if there were no unbalance. The unbalance affects only the realization of the experimental data as we will see in the next section, and it never affects the key formula itself. 

We remark that we have used only the basis independence of the unbalance in our security proof. Thus, it follows that as long as the unbalances are basis-independent, our conclusion holds even if the unbalance of the sending pulses and that of the measurement are unknown and fluctuate in time. It also follows that if we use the so-called 
squash operators \cite{GLLP, Tsurumaru and Tamaki, BML} for Bob's measurement satisfying Eq.~(\ref{Basis Independent Failure}), 
which is shown to exist in \cite{Original}, then one can draw the same conclusion even when one uses any proof technique
of the balanced BB84, including Shor-Preskill type security proof \cite{Lo and Chau,Shor and Preskill}, where 
Eve is regarded as the sender of a basis-independent quantum state to Alice and Bob. 

\section{Simulation}\label{Simulation}
In this section, we simulate the key generation rate by using the typical experimental parameters taken from the Gobby-Yuan-Shields (GYS) experiments \cite{GYS}. In the simulation, we assume the use of infinite number of decoy states to obtain $\gamma_X^{(1)}$ and $e_{Y}^{(1)}$.
We also assume that the bit error only stems from the dark counting of the detectors, and therefore, we ignore the probability of the error stemming from the misalignment and other imperfections of the devices.
In order to ensure the basis-independent detection (Eq.~(\ref{Basis Independent Failure})), we assume that the quantum efficiencies of the two detectors are the same and the inefficiency of the detector can be modeled by a beam splitter preceded by a detector with unit quantum efficiency. Moreover, we assign a random bit value to the double click event.

With all the assumptions, we may have the following experimental data.
\begin{align}
\begin{split}
\eta=&\eta_{det} 10^{-\frac{\xi l}{10}}\\
\beta=&\alpha\eta\\
\gamma_X=&[1-(1-p_{d})e^{-\kappa \beta}](1-p_{d})\\&+(1-p_{d})e^{-\kappa \beta}p_{d}+[1-(1-p_{d})e^{-\kappa \beta}]p_{d}\\
E_X=&\{(1-p_{d})e^{-\kappa \beta}p_{d}+\frac{1}{2}[1-(1-p_{d})e^{-\kappa \beta}]p_{d}\}/\gamma_X\\
p^{(1)}=&e^{-(1+\kappa)\alpha}\alpha(1+\kappa)\\
\gamma_X^{(1)}=&\{[1-(1-p_{d})(1-\eta \frac{\kappa}{1+\kappa})](1-p_{d})\\&+(1-p_{d})(1-\eta \frac{\kappa}{1+\kappa})p_{d}\\&+[1-(1-p_{d})(1-\eta \frac{\kappa}{1+\kappa})]p_{d}\}p^{(1)}\\
e_Y^{(1)}=&\{(1-p_{d})(1-\eta \frac{\kappa}{1+\kappa})p_{d}\\&+\frac{1}{2}[1-(1-p_{d})(1-\eta \frac{\kappa}{1+\kappa})]p_{d} \}p^{(1)}/\gamma_X^{(1)}
\end{split}
\end{align}
Here, $p_{d}$ is the dark count rate of each of the detector D0 and D1, $\eta$ is the overall transmission rate, 
$\eta_{det}$ is the quantum efficiency of the detectors D1 and D2, $\xi$ is the channel transmission rate, and
$l$ is the distance between Alice and Bob.

We also simulate the case of the hardware fix scenario.
In this case, some values of the parameters change as follows.

\begin{align}
\begin{split}
\gamma_X=&[1-(1-p_{d})e^{-\kappa^2 \beta}](1-p_{d})\\&+(1-p_{d})e^{-\kappa^2 \beta}p_{d}+[1-(1-p_{d})e^{-\kappa^2 \beta}]p_{d}\\
E_X=&\{(1-p_{d})e^{-\kappa^2 \beta}p_{d}+\frac{1}{2}[1-(1-p_{d})e^{-\kappa^2 \beta}]p_{d}\}/\gamma_X\\
p^{(1)}=&e^{-2\kappa \alpha}2\kappa \alpha\\
\gamma_X^{(1)}=&\{[1-(1-p_{d})(1-\frac{\eta \kappa}{2})](1-p_{d})\\&+(1-p_{d})(1-\frac{\eta \kappa}{2})p_{d}\\&+[1-(1-p_{d})(1-\frac{\eta \kappa}{2})]p_{d}\}p^{(0)}\\
e_Y^{(1)}=&\{(1-p_{d})(1-\frac{\eta \kappa}{2})p_{d}\\&+\frac{1}{2}[1-(1-p_{d})(1-\frac{\eta \kappa}{2})]p_{d}\}p^{(1)}/\gamma_X^{(1)}
\end{split}
\end{align}

\begin{figure}[bht]
\begin{center}
 \includegraphics[scale=0.90]{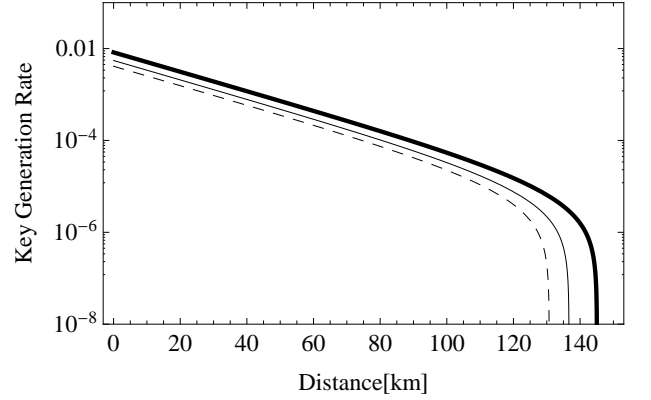}
\end{center}
 \caption{
Key generation rates of the unbalanced BB84. The thick line is a ideal phase encoded BB84 ($\kappa=1$), the thin line is the unbalanced BB84 ($\kappa=1/2$), and dashed line is the BB84 with the hardware fix ($\kappa=1/2$).
}
 \label{KGRGYS}
\end{figure}

\begin{figure}[bth]
\begin{center}
 \includegraphics[scale=0.90]{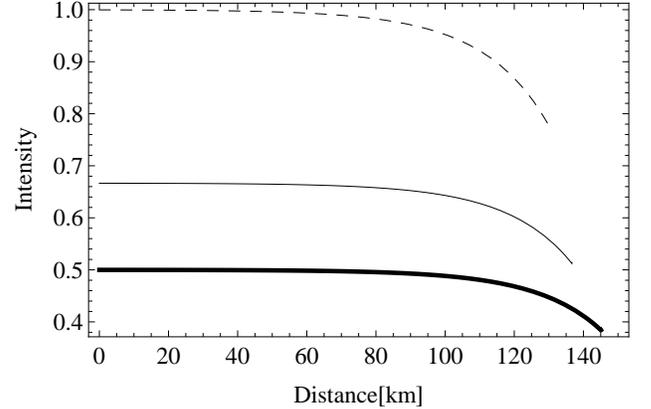}
\end{center}
 \caption{
Optimal intensity $\alpha$ that results in each of the optimal key generation rate in FIG. \ref{KGRGYS}.
}
 \label{AGYS}
\end{figure}

We take the following parameters from the GYS experiments \cite{GYS}: $f(E_X)=1.22$, $p_{d}=8.5\times 10^{-7}$, $\xi=0.21$[dB/km], and $\eta_{det}=0.045$.
The result of the simulation is shown in FIG. \ref{KGRGYS} where we have optimized the intensity of the pulse ($\alpha$), and the optimum intensity is depicted in FIG. \ref{AGYS}. We obtain almost the same transmission distances as those in \cite{Original}. 
Furthermore, we can confirm that the hardware fix scenario causes the decrease in the key generation rate (the same tendency is also obtained by the work by Ferenczi, {\it et. al.} \cite{Original}). This decrease is due to the additional loss in the hardware fix scenario. 

\section{Summary and Discussion}\label{summary}
In summary, we have proved the unconditional security of the unbalanced BB84 protocol. For the security proof, we have considered the virtual protocol that is equivalent to the unbalanced BB84 protocol. In the proof, we have confirmed that the single photon part of the virtual protocol is basis-independent or this part is constructed by the preparation of the Bell state followed by each side of the basis-independent filtering operations, and each side of the two outcome measurements. Thanks to the basis-independence in the virtual protocol, we have concluded that we can apply the method of security proofs of the balanced BB84 \cite{Lo and Chau, Shor and Preskill, Koashi}. Therefore, we can conduct the data processing for the key distillation as if there were no unbalance and the unbalance has influence only through the realization of the experimental data. We note that a natural consequence of our security proof is that as long as the unbalances are basis-independent, our conclusion holds even if the unbalance of the sending pulses and that of the measurement are unknown and fluctuate in time.

Finally, by the simulation, we have also compared the key generation rates of the unbalanced BB84 and the BB84 with the hardware fix, and confirmed that the hardware fix scenario causes the decrease 
in the key generation rate and the transmission distance.

\section{Acknowledgement}
We thank Koji Azuma, Go Kato, William J. Munro, Norbert L\"utkenhaus, and Agnes Ferenczi for valuable discussions and comments. This research is in part supported by the project ``Secure photonic network technology'' as part of ``The project UQCC'' by the National Institute of Information and Communications Technology (NICT) of Japan, in part by the Japan Society for the Promotion of Science (JSPS) through its Funding Program for World-Leading Innovative R$\&$D on Science and Technology (FIRST Program)".


\begin{thebibliography}{99}
\bibitem{comment} For the simplicity, we neglect the contribution of the vacuum emission part to the key generation \cite{KoashiPractical, Lo-nothing} and the noisy processing \cite{noisy} to increase the rate. Since the vacuum emission part is trivially basis-independent, direct application of our proof shows the positive contribution in the unbalanced BB84.
\end{thebibliography}
\end{document}